\documentclass{article}
\usepackage[T1]{fontenc} 
\usepackage[utf8]{inputenc} 
\usepackage{ismir,amsmath,cite,url}
\usepackage{graphicx}
\usepackage{color}
\def\lbd{}	

\usepackage{lineno}

\title{Music Score Expansion with Variable-Length Infilling}

\multauthor
{Chih-Pin Tan$^1$ \hspace{1cm} Chin-Jui Chang$^2$ \hspace{1cm} Alvin W.Y. Su$^1$ \hspace{1cm} Yi-Hsuan Yang$^{2,3}$} {\\
$^1$ Department of Computer Science, National Cheng Kung University\\
$^2$ Research Center for IT Innovation, Academia Sinica\\
$^3$ Yating Music Team, Taiwan AI Labs\\
{\tt\small p76091551@gs.ncku.edu.tw, yhyang@ailabs.tw}
}

\def\authorname{C.P. Tan, C.J. Chang, A. Su, and Y.H. Yang}

\begin{document}

\maketitle
\begin{abstract}
In this paper, we investigate using the 
\textit{variable-length infilling (VLI)} model \cite{chang2021variable}, which is originally proposed to infill missing segments,
to ``prolong''
existing musical segments at musical boundaries. Specifically, as a case study, we expand 20 musical segments from 12 bars to 16 bars, and examine the degree to which the VLI model preserves musical boundaries in the expanded results using a few objective metrics, including the \textit{Register Histogram Similarity} we newly propose. The results show that the VLI model has the potential to address the expansion task.
\end{abstract}

\section{Introduction}
\begin{figure}
 \centerline{
 \includegraphics[width=0.87\columnwidth]{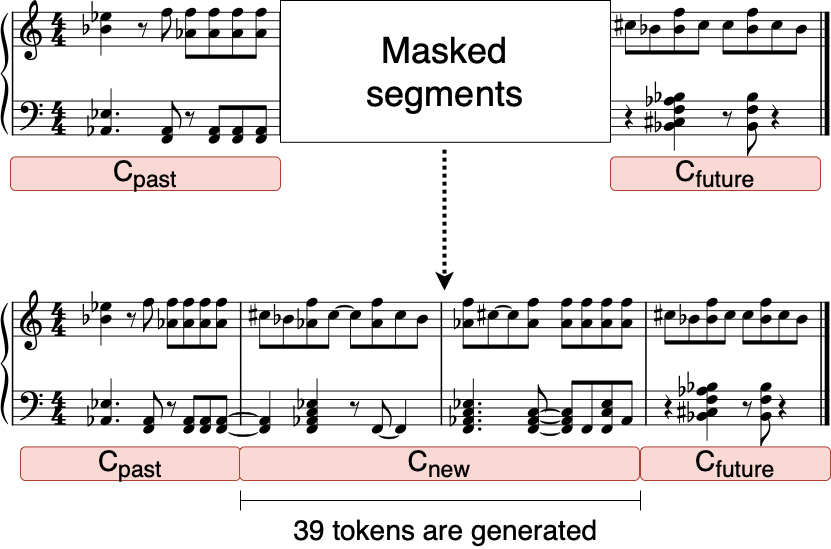}}
 \caption{An illustration of how the variable-length infilling model \cite{chang2021variable} works, where the model is asked to generate 2-bar musical content (i.e., the ``context gap'' is two bars). We do not explicitly consider key in this work.}
 \label{fig:vli}
 \vspace{-3mm}
\end{figure}

\textit{Music score infilling} 
is an instance of automatic symbolic music generation tasks. Given two \emph{disconnected} musical segments, an infilling model aims to ``fill the gap'' between them by generating novel content, as depicted in \figref{fig:vli}. We refer to the two given segments as the \emph{past context} $C_\text{past}$ and the \emph{future context} $C_\text{future}$, respectively,
and the generated one as the \emph{infilled segment} $C_\text{new}$.
We assume that these segments are all represented by sequences of event tokens such as note-on and note-duration \cite{hsiao2021compound}.

Very recently, we
proposed a \textit{variable-length infilling (VLI)} model \cite{chang2021variable} 
that accepts variable-length ``context gaps'' and generates variable-length musical contents.
Specifically, at inference time, the input needed by the VLI model is composed of only the two token sequences $C_\text{past}$, $C_\text{future}$, and \emph{the number of bars} between $C_\text{past}$ and $C_\text{future}$ (i.e., the context gap). 
VLI decides on its own the number of tokens to be filled to the designated gap between the given contexts.
Moreover, a single VLI model can deal with different context gaps (instead of needing one model per context gap).
In VLI, the context gap is expected to be a non-zero and positive integer.

Interestingly, we later realize that VLI also holds the potential to address a highly relevant, yet much less explored task, called \textbf{music score expansion}.
Given a sequence of music notes $\{n_{1}, n_{2},..., n_{L}\}$, this task entails increasing the length of the sequence by \emph{inserting} novel content at one or multiple locations of the sequence.
Without loss of generalizability, we assume that we are about to insert content at only one location $p$, where $1<p<L$. 
We can then use VLI to address this task by treating $\{n_{1}, n_{2},..., n_{p}\}$ as $C_\text{past}$, $\{n_{p+1}, n_{p+2},..., n_{L}\}$ as $C_\text{future}$, and setting the desired context gap to a non-zero integer.
In other words, while the $C_\text{past}$ and $C_\text{future}$ are originally \emph{connected} with zero gap in between, the  idea here is to split them apart somewhere, create artificial gap between then, and use VLI to fill the gap, to prolong the sequence as a result.\footnote{We can also ``shorten'' a sequence by removing a sub-sequence and use VLI to create a novel content that is shorter than the removed sub-sequence to connect the gap smoothly; we leave this as a future work.}

In this paper, we present preliminary experiments exploring such a novel use case of VLI, treating music score expansion as a sub-task of score infilling. 

\begin{figure}
 \centerline{
 \includegraphics[width=1.05\columnwidth]{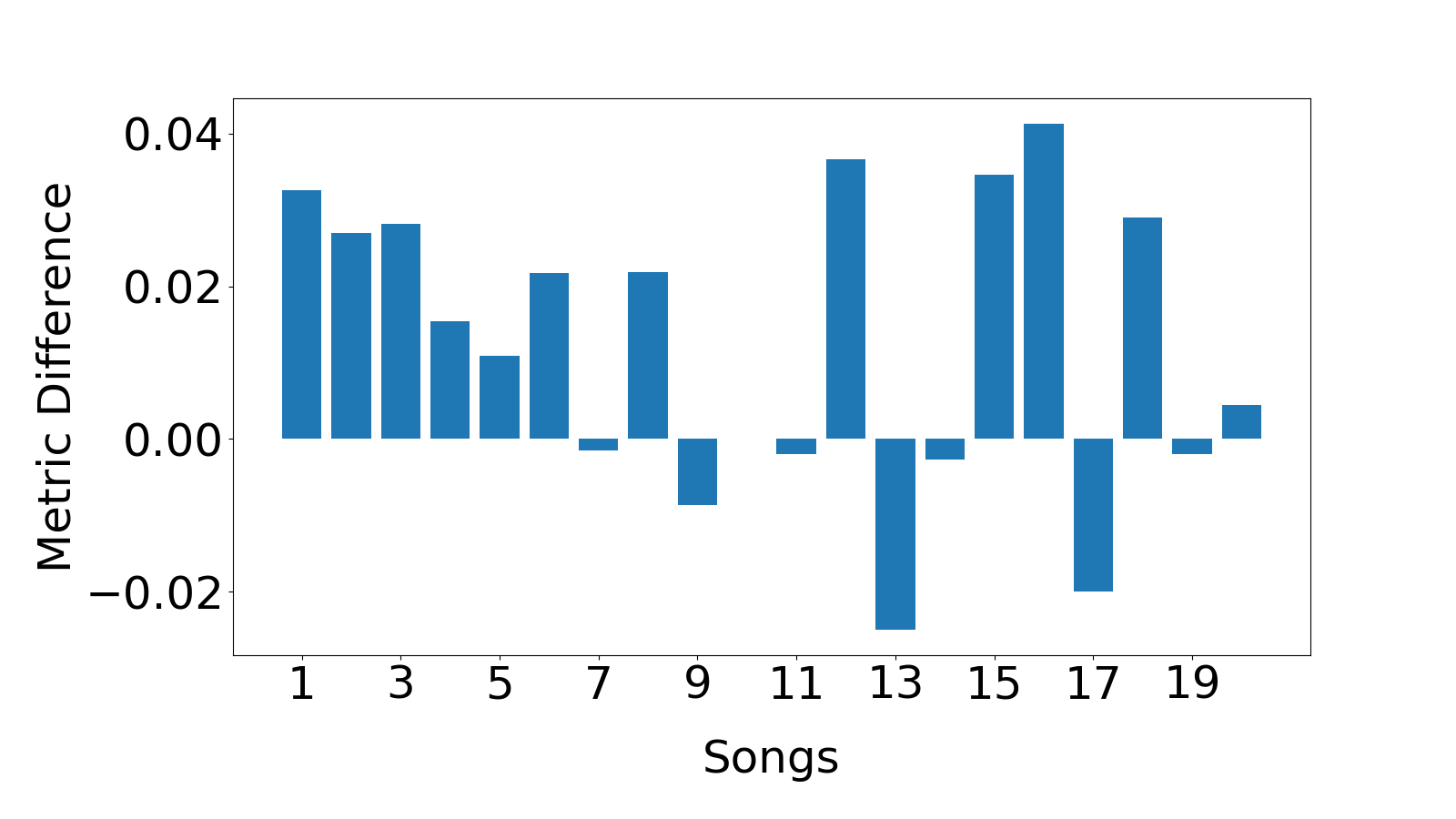}}
 \caption{The result of subtracting the grooving pattern similarity of ($C_\text{past}$, $C_\text{new}$) from that of ($C_\text{new}$, $C_\text{future}$).}
 \label{fig:gs}
\end{figure}

\vspace{-2mm}
\section{Experiment}
While there are many ways to choose the location $p$ for content insertion,
in our work, we utilize the VLI model to expand short-length musical segments at 
\textit{musical boundaries}. A musical boundary is viewed as the edge of two musical groups \cite{lerdahl1996generative}, which is often accompanied by changes in rhythm, metre and pitch patterns. It should be expected in general that there is still a musical boundary after expansion. This is the focus of our experiment.

Specifically, we choose 20 pieces of 12-bar musical segments from the AILabs-Pop1k7 dataset \cite{hsiao2021compound}, which are 
MIDI transcriptions of polyphonic pop piano performances. 
The detailed settings of the VLI model is identical to our previous work \cite{chang2021variable}. 
For each segment, we use pre-trained VLI to insert 4-bar novel content at a manually-chosen boundary which is located at the edge of two \emph{musical phrases}. MIDI files of all the original and expanded segments can be found at the demo website.\footnote{https://tanchihpin0517.github.io/variable-length-piano-expansion/}

\section{Evaluation}
We use two metrics in our evaluation. 
We hypothesize that, if there is a sudden change in \emph{rhythm} patterns at $p$ going from $C_\text{past}$ to $C_\text{future}$, such a sudden change should be preserved within the generated $C_\text{new}$. 
To quantify this, we use the \textit{grooving pattern similarity} ($\mathcal{GS}$) proposed in \cite{wu2020jazz} to compute the similarity in rhythm between $\{C_\text{past}, C_\text{new}\}$ (denoted as $\mathcal{GS}_{1}$) and between $\{C_\text{new}, C_\text{future}\}$ ($\mathcal{GS}_{2}$), respectively, and then compare the scores by subtracting the former from the latter (i.e., $\mathcal{GS}_{2}-\mathcal{GS}_{1}$).
A positive subtraction result indicates that the rhythm pattern of $C_\text{new}$ is closer to that of $C_\text{future}$, otherwise to $C_\text{past}$.
Besides, when there is a boundary, we expect the absolute value of the subtraction result to be a large number, assuming that only $\mathcal{GS}_{1}$ or $\mathcal{GS}_{2}$ would take a large value, but not both.

Across boundaries, there will also be changes in \emph{pitch} patterns due to, e.g., splitting a complete chord into arpeggios or shifting melody line to another octave or key. 
To reflect these, we introduce 
a new
metric called \textit{register histogram similarity} ($\mathcal{RHS}$).
Given a sequence of notes, we count the number of notes in each octave and construct a 7-dimensional histogram $\vec{h}$ from \texttt{C1} to \texttt{C7}. Given the histograms $\vec{h}_{1},\vec{h}_{2}$ of two segments (e.g., $\{C_\text{new}, C_\text{future}\}$), we evaluate their similarity 
by their negative cross entropy (which lies in $(-\infty,0]$, the closer to zero the more similar):
\begin{equation}\label{phce}
\mathcal{RHS}(\vec{h}_{1},\vec{h}_{2})=\sum\nolimits_{i=0}^{7}h_{1,i}\,\log_{2}(h_{2,i})\,.
\end{equation}
We similarly quantify the degree of changes in pitch by computing the $\mathcal{RHS}$ between $\{C_\text{past}, C_\text{new}\}$  and between $\{C_\text{new}, C_\text{future}\}$, respectively, and then subtracting them.


\begin{figure}
 \centerline{
 \includegraphics[width=1.05\columnwidth]{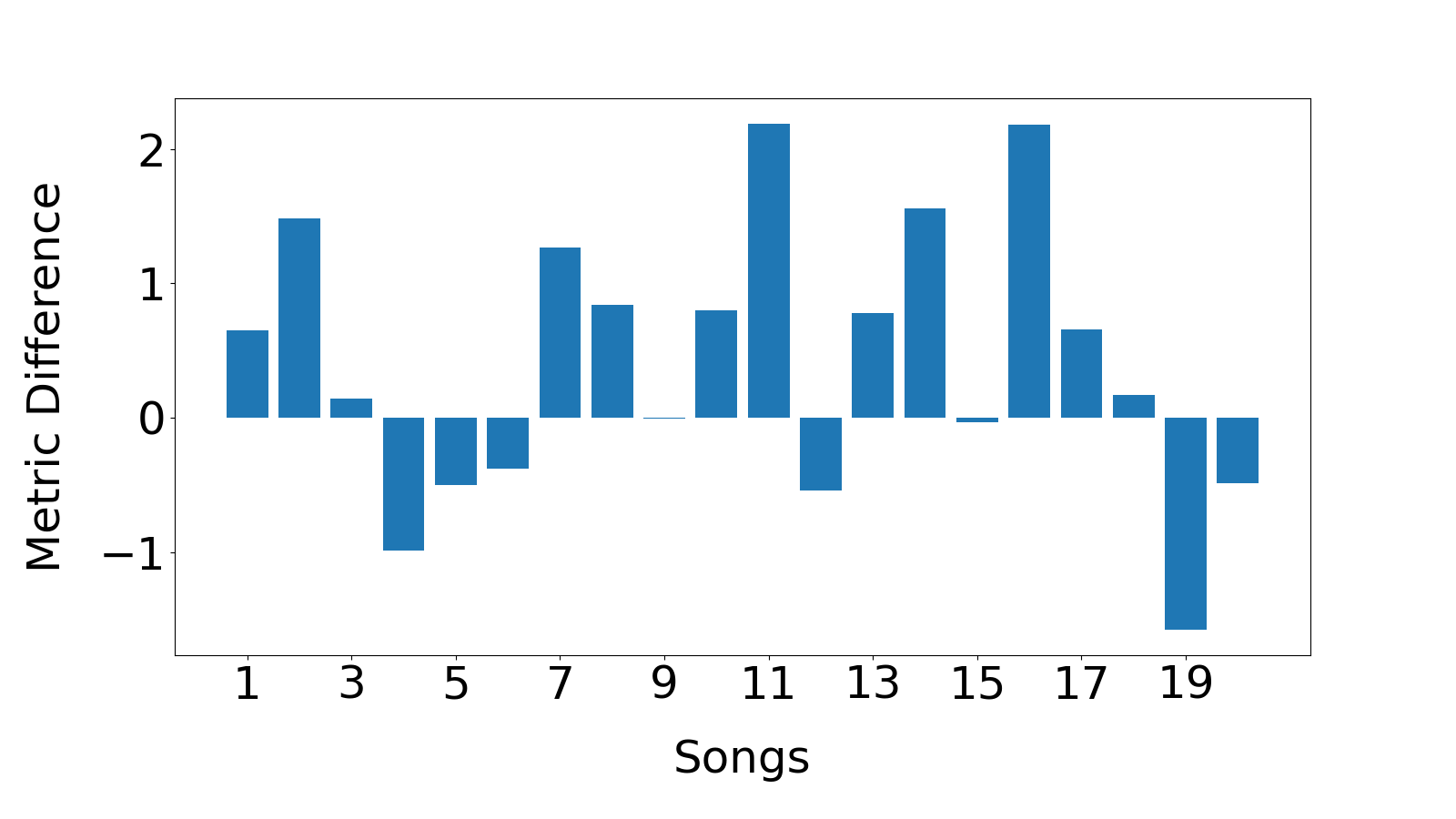}}
 \caption{The result of subtracting the register histogram similarity of ($C_\text{past}$, $C_\text{new}$) from that of ($C_\text{new}$, $C_\text{future}$)}
 \label{fig:rhce}
\end{figure}
\par

Figures \ref{fig:gs} and \ref{fig:rhce} show the result
in $\mathcal{GS}$ and $\mathcal{RHS}$, respectively.
We observe that the VLI model tends to keep the boundary in two ways: i) making $C_\text{new}$ an extended content of $C_\text{past}$ or $C_\text{future}$ by imitating one of them, or ii) making $C_\text{new}$ an independent segment which plays the role of a bridge connecting $C_\text{past}$ and $C_\text{future}$.
The boundary is often preserved well if the absolute subtraction result in $\mathcal{GS}$ or $\mathcal{RHS}$ is large, but not vice versa---informal listening shows that VLI is able to preserve the boundary even when the absolute subtraction results in both $\mathcal{GS}$ and $\mathcal{RHS}$ are low.
We also find that VLI tends to make $C_\text{new}$ more similar to $C_\text{future}$ than to $C_\text{past}$, possibly because it is harder to extend a segment which has already a proper sense of ending, which is often the case for $C_\text{past}$ in pop music.

%
\section{Conclusion}
In this paper, we have shown that the VLI model performs well for music score expansion in certain conditions. However, we currently use human-annotated boundaries for generation tasks. Selecting a target place for expanding is still an ongoing research topic. We aim to integrate music-structure analysis into the model and make it capable to decide where to insert content. Moreover, the metrics in our work are rather simplistic, describing the properties of boundaries fairly roughly. Finding other proper metrics is also one of our next steps.


\begin{thebibliography}{1}
\providecommand{\url}[1]{#1}
\csname url@samestyle\endcsname
\providecommand{\newblock}{\relax}
\providecommand{\bibinfo}[2]{#2}
\providecommand{\BIBentrySTDinterwordspacing}{\spaceskip=0pt\relax}
\providecommand{\BIBentryALTinterwordstretchfactor}{4}
\providecommand{\BIBentryALTinterwordspacing}{\spaceskip=\fontdimen2\font plus
\BIBentryALTinterwordstretchfactor\fontdimen3\font minus
  \fontdimen4\font\relax}
\providecommand{\BIBforeignlanguage}[2]{{%
\expandafter\ifx\csname l@#1\endcsname\relax
\typeout{** WARNING: IEEEtran.bst: No hyphenation pattern has been}%
\typeout{** loaded for the language `#1'. Using the pattern for}%
\typeout{** the default language instead.}%
\else
\language=\csname l@#1\endcsname
\fi
#2}}
\providecommand{\BIBdecl}{\relax}
\BIBdecl

\bibitem{chang2021variable}
C.-J. Chang \emph{et~al.}, ``Variable-length music score infilling via {XLNet}
  and musically specialized positional encoding,'' in \emph{Proc. ISMIR}, 2021.

\bibitem{hsiao2021compound}
W.-Y. Hsiao \emph{et~al.}, ``{Compound Word Transformer}: Learning to compose
  full-song music over dynamic directed hypergraphs,'' in \emph{Proc. AAAI},
  2021.

\bibitem{lerdahl1996generative}
F.~Lerdahl and R.~S. Jackendoff, \emph{A Generative Theory of Tonal
  Music}.\hskip 1em plus 0.5em minus 0.4em\relax MIT Press, 1996.

\bibitem{wu2020jazz}
S.-L. Wu and Y.-H. Yang, ``The {Jazz Transformer} on the front line: Exploring
  the shortcomings of {AI}-composed music through quantitative measures,'' in
  \emph{Proc. ISMIR}, 2020.

\end{thebibliography}

%
%
%
%
%

\end{document}